\documentclass[onecolumn,superscriptaddress,preprintnumbers,nofootinbib,notitlepage,cite,showpacs]{revtex4}

\usepackage{bm} 
\usepackage{simplewick}
\usepackage{graphics,graphicx}
\usepackage{amsmath,amssymb}
\usepackage{color}
\usepackage{array,mathtools}
\usepackage{booktabs,multirow}
\usepackage{bbold}
\usepackage[normalem]{ulem}
\usepackage[colorinlistoftodos,prependcaption,textsize=tiny]{todonotes}
\usepackage{braket}

\newcommand{\bea}{\begin{eqnarray}}
\newcommand{\eea}{\end{eqnarray}}

\newcommand{\beq}{\begin{equation}}

\newcommand{\fm}{\rm fm}
\newcommand{\eeq}{\end{equation}}

\newcommand{\MeV}{{\rm MeV}}
\newcommand{\itGamma}{\mathit{\Gamma}}
\newcommand{\vecb}[1]{\boldsymbol{#1}}
\newcommand{\matrixb}[1]{\boldsymbol{#1}}
\newcommand{\smex}{\hspace{-0.05em}} 
\newcommand{\PP}{P \! P} \newcommand{\PPv}{\PP \smex v} \newcommand{\vPPv}{v \smex \PP \smex v} \newcommand{\APv}{A \! P \smex v}
\newcommand{\VV}{V \! V} \newcommand{\VVv}{\VV \! v} \newcommand{\vVVv}{v \smex \VV \! v} \newcommand{\TVv}{T\smex V \! v}
\newcommand{\myfrac}[2]{\genfrac{}{}{}{}{\raisebox{-1.5pt}{$#1$}}{\raisebox{1pt}{$#2$}}}
\newcommand{\arcosh}{\mathrm{arcosh}}
\newcommand{\half}{\frac{1}{2}}
\newcommand{\DeltaE}{\Delta_{E}}
\newcommand{\PV}{{\{P,V\}}}

\definecolor{nred}{RGB}  {240, 50,  0}
\definecolor{nblue}{RGB} {  0, 50,200}
\definecolor{ngreen}{RGB}{ 50,200,  0}

\makeatother

\begin{document}

\preprint{\tt LPT-Orsay-18-23}
\preprint{\tt MS-TP-18-10}

\vspace*{30mm}

\title{Leptonic $\mbox{\boldmath$D_s$}$ decays in two-flavour lattice QCD}

\author{Beno\^it~Blossier}
\affiliation{Laboratoire de Physique Th\'eorique\footnote[1]{Unit\'e Mixte de Recherche 8627 du Centre National de la Recherche Scientifique}, CNRS, Univ. Paris-Sud et Universit\'e Paris-Saclay,  B\^atiment 210,  91405 Orsay Cedex, France}
\author{Jochen Heitger}
\affiliation{Institut~f\"ur~Theoretische~Physik, Westf\"alische Wilhelms-Universit\"at~M\"unster, Wilhelm-Klemm-Str.~9, 48149~M\"unster, Germany}
\author{Matthias Post}
\affiliation{Institut~f\"ur~Theoretische~Physik, Westf\"alische Wilhelms-Universit\"at~M\"unster, Wilhelm-Klemm-Str.~9, 48149~M\"unster, Germany}

\begin{abstract}
We report on a two-flavour lattice QCD study of the $D_s$ and $D^*_s$ leptonic decays parameterized by the decay constants $f_{D_s}$ and $f_{D^*_s}$. As the phenomenology in the $D_s$ sector seems very promising in the next years with the experiments LHCb and Belle~II, it is worth putting a big effort in lattice computations regarding its non-perturbative QCD contributions. Before examining more challenging processes such as hadron-hadron transitions, a natural first step is to address some basic aspects in the context of leptonic decays, where systematic uncertainties from excited state contaminations and cutoff effects in the computation of charmed meson decay matrix elements can be investigated in a more straightforward setting. 
\end{abstract}

\pacs{11.15.Ha, 12.38.Gc, 13.20.Fc, 14.65.Dw}

\maketitle

\section{Introduction}

After 60 years of joint effort by theoretical and experimental communities in particle physics, the Standard Model (SM) offers a complete picture of fundamental interactions up to the electroweak scale. In the quark sector, for instance, which is spread in 3 families, charged weak decays are mediated by a left-handed current with the exchange of a $W$ boson, flavour changing neutral currents are forbidden at tree-level by virtue of the so-called Glashow-Iliopoulos-Maiani (GIM) mechanism \cite{GlashowGM}, and CP violation shows up because the Jarlskog invariant $J$ is different from zero \cite{JarlskogHT}. On the electroweak side, the presence of a scalar field with a non-zero vacuum expectation value (VEV) induces a spontaneous breaking of the $SU(2)_{\rm W} \times U(1)_{\rm Y}$ electroweak symmetry into $U(1)_{\rm EM}$, which manifests in the Higgs mechanism \cite{EnglertET, HiggsPJ, GuralnikEU}: Starting from a complex iso-doublet of scalar fields, 3 degrees of freedom are absorbed to give masses to the $W$ and $Z$ bosons such that a single field remains, the Higgs boson $H$. With a mass of 125.09(24) GeV (as measured along the data taking of LHC Run 1 \cite{AadZHL}) and a VEV of 246 GeV, its coupling to vector bosons is quadratic in their masses, while its coupling to quarks is linear in their masses. 

A well-known issue with the SM Higgs is, however, that the quartic term in the Higgs Lagrangian generates for the Higgs mass $m_H$ a quadratic divergence with the hard scale of the theory, related to the so-called hierarchy problem \cite{WilsonAG, tHooftRAT}. There are several New Physics scenarios that are supposed to cure this caveat (amongst the others) of the SM.
One class of proposed scenarios beyond the SM is characterized by a minimal extension of the Higgs sector. They contain 2 complex scalar iso-doublets $\Phi_1$ and $\Phi_2$, which after the spontaneous breaking of the electroweak symmetry lead to 2 charged particles $H^\pm$, 2 CP-even particles $h$ (an SM-like Higgs) and $H$ plus 1 CP-odd particle $A$ (see \cite{DjouadiGJ, Gunion} and references therein for nice reviews). In these scenarios, quarks are coupled to a charged Higgs through a right-handed current. It is particularly this feature that has received a lot of attention recently, because several tests of lepton-flavour universality have shown some hints of an anomaly with respect to SM expectations, especially for the ratios $R_{D^{(*)}}\equiv \frac{\Gamma(B\to D^{(*)} \tau \nu_\tau)}{\Gamma(B \to D^{(*)} \ell \nu_\ell)}, \ell=e,\mu$ \cite{LeesXJ, HuschleRGA, AaijYRA}: Semi-leptonic decays with a $\tau$ lepton in the final state can have a non-SM contribution from the exchange of a right-handed current that is not helicity-suppressed by the mass of the charged lepton. In view of the highly promising perspective that Belle~II uses a part of its integrated luminosity to run at the energy of $\Upsilon (5S)$ in order to accumulate $B_s$ pairs, it then might be very valuable to investigate equivalent ratios $R_{D^{(*)}_s}$, where just the spectator quark in the aforementioned processes is changed. Therefore, on the theory side, the non-perturbative hadronic properties of the $B_s$, $D_s$ and $D^*_s$ mesons involved, which are accessible through lattice QCD, have to be under very good control. 

This paper represents the very first step in this program, and it reports on an estimate by $N_f=2$ lattice QCD of the ratio of charm-strange leptonic decay constants in the vector and pseudoscalar channels, $f_{D^*_s}/f_{D_s}$, which quantifies spin-breaking effects in heavy-strange mesons. While a large number of results for $f_{D_s}$ is available in the literature \cite{AubinAR, FollanaUV, BlossierBX, DaviesIP, NamekawaWT, DimopoulosGX, BazavovAA, NaIU, CarrascoZTA, ChenHVA, BazavovWGS, YangSEA, CarrascoPOA, Boyle:2017jwu}, there are so far only a few of them for $f_{D^*_s}$ \cite{BecirevicTI, DonaldSRA, LubiczASP}; moreover, this study serves as an opportunity to learn how to identify and alleviate systematic effects such as contaminations from excited states in correlation functions and potentially large lattice artifacts, which are typically encountered in simulations of $D_s$ meson systems.

\section{Lattice computation}
 
\renewcommand{\arraystretch}{1.2}
\setlength{\tabcolsep}{8pt}
\begin{table}[t]
\begin{center}
\begin{tabular}{lcc@{\hskip 02em}c@{\hskip 02em}c@{\hskip 01em}c@{\hskip 01em}c@{\hskip 01em}c@{\hskip 01em}c@{\hskip 01em}c}
\hline
\hline
	\toprule
	id	&	$\quad\beta\quad$	&	$(L/a)^3\times (T/a)$ 		&	$\kappa_{\rm sea}$		&	$a~[\rm fm]$	&	$m_{\pi}~[\MeV]$	& $Lm_{\pi}$ 	& $\#$ cfgs&$\kappa_s$&$\kappa_c$\\
\hline 
\hline 
	\midrule
	E5	&	5.3		&	$32^3\times64$	& 	$0.13625$	& 	0.0653	  	& 	$439$	&4.7	& $200$&$0.135777$&$0.12724$\\  
	F6	&			& 	$48^3\times96$	&	$0.13635$	& 			& 	$313$	&5	& $120$&$0.135741$&$0.12713$\\    
	F7	&			& 	$48^3\times96$	&	$0.13638$	& 			& 	$268$	&4.3	& $200$&$0.135730$&$0.12713$\\    
	G8	&			& 	$64^3\times128$	&	$0.13642$	& 			& 	$194$	&4.1	& $176$&$0.135705$&$0.12710$\\    
\hline
	\midrule
	N6	&	$5.5$	&	$48^3\times96$	&	$0.13667$	& 	$0.0483$  	& 	$341$	&4	& $192$&$0.136250$&$0.13026$\\	
	O7	&		&	$64^3\times128$	&	$0.13671$	& 	 	& 	$269$	&4.2	& $160$&$0.136243$&$0.13022$ \\ 
	\bottomrule
\hline
\hline 
\end{tabular} 
\end{center}
\caption{Parameters of the two-flavour gauge field configuration ensembles underlying the simulations of this study: Bare coupling $\beta = 6/g_0^2$, lattice resolution, the sea quark's hopping parameter $\kappa_{\rm sea}$, lattice spacing $a$ in physical units, pion mass, number of configurations, as well as the hopping parameters in the valence sector corresponding to (bare) strange and charm quark masses, respectively.}
\label{tab:sims}
\end{table}

\subsection{Lattice set-up and analysis techniques}

This work is based on a subset of the CLS ensembles \cite{CLSweb}, made of $N_f=2$ non-perturbatively $\mathrm{O}(a)$ improved Wilson-Clover fermions \cite{SheikholeslamiIJ, LuscherUG}, together with the plaquette gauge action \cite{WilsonSK} for gluon fields, and generated using either the DD-HMC algorithm \cite{LuscherQA, LuscherRX, LuscherES, Luscherweb} or the MP-HMC algorithm \cite{MarinkovicEG}. We collect our simulation parameters in Table~\ref{tab:sims}. Two values of the lattice spacing, $a_{\beta=5.5}=0.04831(38)\,\fm$ and $a_{\beta=5.3}=0.06531(60)\,\fm$ as determined from a fit in the chiral sector \cite{LottiniRFA}, are considered, with pion masses in the range $[190\,, 440]~\MeV$. The bare (valence) strange quark mass had been tuned by imposing the ratios $m^2_K/f^2_K$ and $m^2_\pi/f^2_K$ to coincide with their physical values, where the scale is set by $f_K$ \cite{Fritzsch:2012wq}. Irrespective of the sea quark mass, the bare mass parameter of the valence charm quark (represented by the hopping parameter $\kappa_c$) was fixed in \cite{HeitgerOAA} through a linear interpolation of $m^2_{D_s}$ in $1/\kappa_c$ to the physical $D_s$ meson mass ($=1968\,\MeV$ \cite{Patrignani:2016xqp}), since this functional form turned out to be best supported by the actual data in the vicinity of the targeted $m^{\rm phys}_{D_s}$. Eventually, statistical errors on each ensemble are estimated from the jackknife procedure\footnote{Since there is enough separation in the Monte-Carlo trajectories between 2 successive measurements, autocorrelation effects can safely be neglected.}, whereas statistical errors on quantities calculated as results of a final joint chiral and continuum limit extrapolation to the physical point are obtained by the following bootstrap-inspired prescription: Create a large set of $N_{\rm event}$ ``event" vectors, the dimension of which is given by the number of CLS ensembles considered in our analysis ($=6$ in the present case), and fill them component-wise with entries randomly chosen from the available sample of $N_{\rm bin[\#id]}$ jackknife-binned data per ensemble (\#{\rm id}=1,...,6). The statistical error of the result of any extrapolating fit is then estimated as the variance over $N_{\rm event}$ such fits, taking these random vectors (which via their very composition can be understood as being drawn from the actual statistical distribution of the contributing raw ensemble data) as inputs.

Two-point correlation functions are evaluated in the standard manner by first expressing them as expectation values of a product of two quark propagators (indicated as $[...]$ in the equation below), where the latter are computed using stochastic sources defined in a randomly chosen timeslice with spin dilution. In addition, the noise has been reduced by applying the one-end trick \cite{FosterWU, McNeileBZ}.
We study the two-point correlation functions
\begin{equation}
C_{\itGamma \itGamma^\prime}(t)= \frac{1}{V} \sum_{\vecb{x},\vecb{y}} \braket{ [\bar{c}\itGamma s](\vecb{y},t) [\bar{s}\gamma_0 \itGamma^\prime \gamma_0 c](\vecb{x},0) },
\end{equation} 
where $V$ is the spatial volume of the lattice, $\langle ... \rangle$ denotes the expectation value over gauge configurations, and the interpolating fields $\bar{c} \Gamma s$ are not necessarily local. Four Gaussian smearing levels for the quark fields $s$ and $c$, including the case of no smearing, are considered to build a $4\times 4$ matrix of correlators, from which we extract all $\mathrm{O}(a)$ improved hadronic quantities relevant here, after analyzing the associated generalized eigenvalue problem (GEVP) \cite{MichaelNE, LuscherCK, BlossierKD}. Solving the GEVP for the pseudoscalar-pseudoscalar and vector-vector matrix of correlators,
\begin{align}
\matrixb{C}_{\PP}(t) \, \vecb{v}^P_n(t,t_0) = \lambda^P_n(t,t_0) \, \matrixb{C}_{\PP}(t_0) \, \vecb{v}^P_n(t,t_0), \\ 
\matrixb{C}_{\VV}(t) \, \vecb{v}^V_n(t,t_0)=\lambda^V_n(t,t_0) \, \matrixb{C}_{\VV}(t_0) \, \vecb{v}^V_n(t,t_0),
\end{align}
we have constructed the corresponding projected correlators (as well as their symmetric counterparts by exchanging operators at the source and at the sink); as local quark bilinears, we here employ the composite fields
\begin{equation}
P=\bar{c}\gamma_5 s, \quad  A_0=\bar{c}\gamma_0\gamma_5 s, \quad V_k=\bar{c}\gamma_k s \quad\text{and}\quad T_{k0}=\bar{c}\gamma_k\gamma_0 s. 
\end{equation} 
These projections, together with their asymptotic behaviour as $t/a \gg 1$, read
\begin{align}
\widetilde{C}_{\vPPv}(t) &= \sum_{i,j} (v^{P}_1)_i(t,t_0) \, C_{P^{(i)} \! P^{(j)}}(t) \, (v^{P}_1)_j(t,t_0)
&\longrightarrow & \;\; \frac{{\cal Z}_{\vPPv}}{am_P} e^{-m_P T/2} \cosh [m_P (T/2 -t)] , \nonumber\\
\widetilde{C}_{\PPv}(t) &= \sum_i C_{P^L \! P^{(i)}}(t) \, (v^{P}_1)_i(t,t_0) 
&\longrightarrow & \;\; \frac{{\cal Z}_{\PPv}}{am_P} e^{-m_P T/2} \cosh [m_P (T/2 -t)] , \nonumber\\
\widetilde{C}_{\APv}(t) &= \sum_i C_{A^L_0 P^{(i)}}(t) \, (v^{P}_1)_i(t,t_0) 
&\longrightarrow & \;\; -\frac{{\cal Z}_{\APv}}{am_P}e^{-m_P T/2} \sinh [m_P (T/2-t )] , 
\label{cfasymptotics_PS}\\[1.0em]
\widetilde{C}_{\vVVv}(t) &= \frac{1}{3} \sum_{i,j,k} (v^{V}_1)_i(t,t_0) \, C_{V^{(i)}_k V^{(j)}_k}(t) \, (v^{V}_1)_j(t,t_0)
&\longrightarrow & \;\; \frac{{\cal Z}_{\vVVv}}{am_V} e^{-m_V T/2} \cosh [m_V (T/2 -t) ] , \nonumber\\
\widetilde{C}_{\VVv}(t) &= \frac{1}{3} \sum_{i,j} C_{V^L_k V^{(i)}_k}(t) \, (v^{V}_1)_i(t,t_0)
&\longrightarrow & \;\; \frac{{\cal Z}_{\VVv}}{am_V} e^{-m_V T/2} \cosh [m_V (T/2 -t) ] , \nonumber\\
\widetilde{C}_{\TVv}(t) &= \frac{1}{3} \sum_{i,k} C_{T^L_{k0} V^{(i)}_k}(t) \, (v^{V}_1)_i(t,t_0) 
&\longrightarrow & \;\; \frac{{\cal Z}_{TV}}{am_V} e^{-m_V T/2} \sinh [m_V (T/2 -t)] ,
\label{cfasymptotics_V}
\end{align}
where the various ${\cal Z}$ stand for the meson-to-vacuum matrix elements of the operators in the respective channels, which arise in the spectral decompositions of the correlation functions. The label ``$L$" refers to a local interpolating field, while the sums over $i$ and $j$ run over the 4 smearing levels.

Upon including the appropriate $\mathrm{O}(a)$ terms in the definitions of the axial and vector currents, viz. 
\begin{align}
A^I_0 &= (1+b_A Z a m_{cs}^{\text{AWI}})(A_0 + c_A \myfrac{a}{2}(\partial_0 \! + \smex \partial^\ast_0)P) ,\\
V^I_k &= (1+b_V Z a m_{cs}^{\text{AWI}})(V_k + c_V \myfrac{a}{2} (\partial_\nu \! + \smex \partial^\ast_\nu) T_{k\nu}) ,
\end{align} 
the forward ($\partial_\nu$) and backward ($\partial^\ast_\nu$) difference operators act on the (foregoing large-$t/a$ asymptotics of the) projected correlators as
\begin{align}
\frac{\partial_0 + \smex \partial^\ast_0}{2} \widetilde{C}_{\PPv}(t) &= - \frac{\sinh(am_P)}{a} \tanh[m_P(T/2-t)] \, \widetilde{C}_{\PPv}(t) ,\\
\frac{\partial_0 + \smex \partial^\ast_0}{2} \widetilde{C}_{\TVv}(t) &= - \frac{\sinh(am_V)}{a} \, \frac{\widetilde{C}_{\TVv}(t)}{\mbox{tanh}[m_P(T/2-t)]} .
\end{align}
Then, together with the asymptotic behaviour of the correlation functions in eqs.~(\ref{cfasymptotics_PS}) and (\ref{cfasymptotics_V}), the lattice expressions for the pseudoscalar- and vector-to-vacuum matrix element of the renormalized axial and vector current, which are proportional to leptonic pseudoscalar and vector meson decay constants of interest, can be split into leading and $\mathrm{O}(a)$ improvement contributions according to:
\begin{align}
& \langle 0|A^R_0|P(\vecb{p}=\vecb{0})\rangle = -f_P m_P = -Z_A(1 + b_A Z a m^{\text{AWI}}_q) m_P f^0_P(1 + f^1_P/f^0_P), \nonumber\\[0.5em]
& \text{with}\quad af^0_P=\frac{1}{am_P} \frac{{\cal Z}_{\APv}}{\sqrt{{\cal Z}_{\vPPv}}}, \quad
af^1_P =  \frac{1}{am_P} c_A \sinh (a m_P) \frac{{\cal Z}_{\PPv}}{\sqrt{{\cal Z}_{\vPPv}}}; 
\label{matrixelement_PS}\\[1.0em]
& \langle 0|V^R_i|V(\vecb{p}=\vecb{0},\lambda)\rangle = \epsilon_i^\lambda f_V m_V = \epsilon_i^\lambda Z_V(1 + b_V Z a m^{\text{AWI}}_q) m_V f^0_V(1 + f^1_V/f^0_V), \nonumber\\[0.5em]
& \text{with}\quad af^0_V=\frac{1}{am_V} \frac{{\cal Z}_{\VVv}}{\sqrt{{\cal Z}_{\vVVv}}}, \quad af^1_V = - \frac{1}{am_V}c_V \sinh (a m_V) \frac{{\cal Z}_{\TVv}}{\sqrt{{\cal Z}_{\vVVv}}},
\label{matrixelement_V}
\end{align}
where $\vecb{p}$ and $\epsilon_i^\lambda$ stand for the spatial meson three-momentum and the vector meson polarization, respectively. 

The renormalization constants $Z_A$ and $Z_V$, which multiply the (time component of the) axial current and the (spatial components of the) vector current, were determined non-perturbatively for two-flavour QCD with $\mathrm{O}(a)$ improved Wilson fermions in \cite{DellaMorteXB, DellaMorteRD}. For the corresponding improvement coefficients $c_A$, $c_V$, $b_A$ and $b_V$ we have used non-perturbative estimates, if available, and perturbative formulas elsewhere \cite{DellaMorteAQE, SintJX}. Moreover, the mass dependent factors in eqs.~(\ref{matrixelement_PS}) and (\ref{matrixelement_V}) involve the matching coefficient $Z$ between the average of bare quark masses defined via the axial Ward identity (also called PCAC quark mass) and its counterpart defined through the vector Ward identity (also called subtracted quark mass). On the lattice, and specifically written for the charm-strange sector considered in this work, these two definitions of $m_{cs}\equiv (m_c+m_s)/2$ translate into
\begin{equation}
m_{cs}^\text{AWI}(t)=\frac{\half(\partial_0 +\partial^*_0) C_{A^L_0P^L}(t) +a c_A \partial_0 \partial^*_0 C_{P^LP^L}(t)}{2C_{P^LP^L}(t)} \quad \text{and} \quad am_{cs}^\text{VWI}=\frac{1}{2} \left(\frac{1}{2\kappa_c}+\frac{1}{2\kappa_s} - \frac{1}{\kappa_{cr}}\right);
\label{mpcac}
\end{equation}
in practice, the local mass $m_{cs}^\text{AWI}(t)$ exhibits an extended plateau over the timeslices sufficiently far from the boundaries of the lattice and therefore can be accurately determined with confidence as the plateau average over central timeslices. Coming back to the factor $Z$ in the present $N_f=2$ case, non-perturbative numbers for it were taken from \cite{Fritzsch:2010aw,Fritzsch:2012wq}. The uncertainties of the non-perturbative values of those various $c$- and $b$-coefficients and $Z$-factors, as far as they are available from the literature, were incorporated in the subsequent analysis, adding them in quadrature with the independent statistical errors of our quantities in question. Thereby, these effects are properly propagated into the final results, yet their contribution is subdominant compared to the combined statistical and lattice spacing errors (from the scale setting) discussed below.

\subsection{Results}

\begin{figure}
 \begin{minipage}[c]{0.49\linewidth}
	\centering 
	\includegraphics*[width=1.0\linewidth]{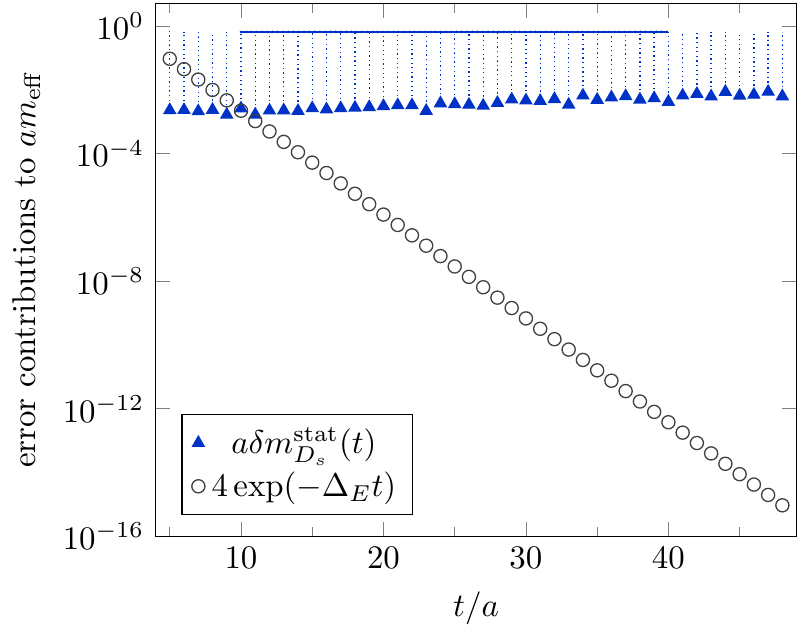}
 \end{minipage}
 \begin{minipage}[c]{0.49\linewidth}
	\centering 
	\includegraphics*[width=0.975\linewidth]{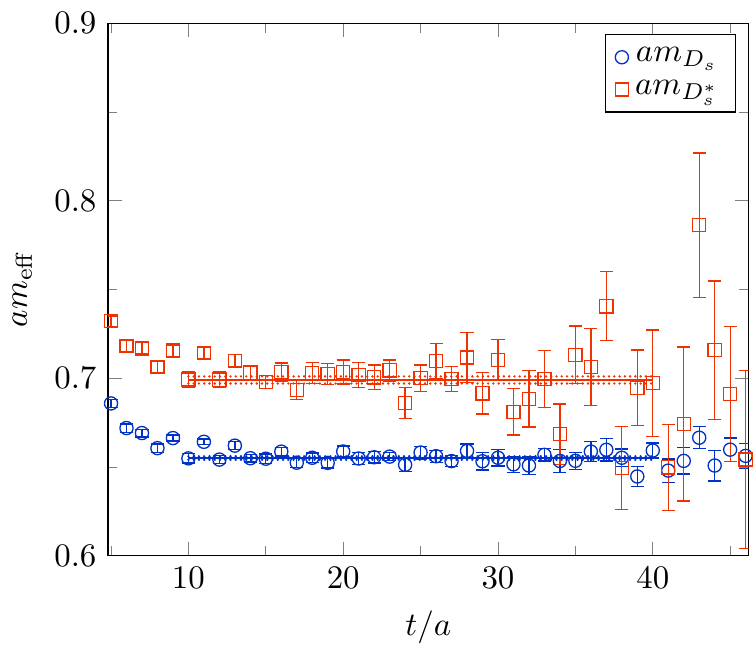}
 \end{minipage}
 \caption{Left panel: 
Statistical error and four times the systematic error contribution $a\Delta m^{\rm sys}$ to the effective mass $am_{D_s}$ from a $4\times 4$ GEVP for the lattice ensemble F7, where the latter ---~due to excited state contaminations~--- is safely modeled as $a\Delta m^{\rm sys}(t)=\exp[-\DeltaE t]$ with $\Delta E=E_4-E_1\sim 2\,{\rm GeV}$. The ground-state mass is then determined over a fit interval (actually indicated by the blue horizontal line for illustrative purposes only), which starts at a $t_{\rm min}$ such that $\delta m^{\rm stat}(t_{\rm min})>4\Delta m^{\rm sys}(t_{\rm min})$ and hence the systematic effect from residual excited state contaminations in this region is very well covered by the statistical uncertainty. Right panel: Effective masses $am_{D_s}$ and $am_{D^*_s}$ extracted from a $4\times 4$ GEVP, again for dataset F7, where also the plateaux over the chosen fit interval is displayed.}
\label{fig1}
\end{figure}

As a starting point of our analysis to extract the charm-strange pseudoscalar and vector meson masses and decay constants along the lines outlined in the last subsection, we have fitted the projected correlators $\widetilde{C}$ within a time range $[t_{\rm min}, t_{\rm max}]$ such that the statistical error $\delta m^{\rm stat}(t_{\rm min})$ on the ground-state effective mass $E_1$ is significantly larger than the systematic error caused by contaminations from higher, excited states. As an estimate of the latter, we advocate a mass gap of 
\begin{equation}
a\Delta m^{\rm sys}(t_{\rm min}) \equiv \exp[-\DeltaE t_{\rm min}] \quad\text{with}\quad \DeltaE = E_4 - E_1  \sim 2\,{\rm GeV}.
\end{equation} 
To be on the conservative side, we always imposed $\delta m^{\rm stat}(t_{\rm min})> 4\Delta m^{\rm sys}(t_{\rm min})$. This is also visualized semi-logarithmically in the left panel of Fig.~\ref{fig1} for a typical case, which demonstrates that, by virtue of our prescription to select the fitting interval for the ground-state extraction, the systematic effect from residual excited state contaminations within this interval is negligible by contrast with the statistical errors. Despite having solved a $4\times 4$ GEVP, we notice that the third excited state is not well under control, because huge statistical errors on the corresponding effective mass are encountered. Nevertheless, based on the overall landscape of resulting energies from the low-lying states, we find our above guesstimate $\DeltaE \sim 2\,{\rm GeV}$ to be safe. The upper bound $t_{\rm max}$ of the particular fit interval per ensemble was fixed by an individual visual inspection of the quality of the effective mass plateaux; its actual influence on the final numbers is insignificant though, as long as one stays within the plateau region. Since it can be assumed (and was numerically confirmed on a few representative datasets) that $\DeltaE^{P} \sim \DeltaE^{V}$ holds to a good approximation for the mass gaps in the pseudoscalar and vector channels at fixed lattice spacing and pion mass, identical fit intervals were chosen for pseudoscalar and vector meson states. Lastly, as for the timeslice parameter $t_0$ in the GEVP analysis (see eqs.~(\ref{cfasymptotics_PS}) and (\ref{cfasymptotics_V})), we have stuck to $t_0=3a$ at $\beta=5.3$ and $t_0=5a$ at $\beta=5.5$, after we had observed that the resulting energies and matrix elements do not change appreciably and only the statistical errors grow.

The right panel of Fig.~\ref{fig1} illustrates the time dependence of the effective masses of the $D_s$ and $D^*_s$ mesons obtained from the gauge configuration ensemble F7. They were calculated by the formulas
\begin{equation}
am^{\PV,{\rm eff}}(t) = \arcosh \left(\frac{\lambda^\PV_1(t+a,t_0)+\lambda^\PV_1(t-a,t_0)}{2\lambda^\PV_1(t,t_0)}\right) .
\end{equation}
In Table \ref{tab:rawdata} of the appendix we specify all masses and decay constants as they result from the application of the GEVP analysis, in conjunction with eqs.~(\ref{cfasymptotics_PS})~--~(\ref{matrixelement_V}), to the CLS ensembles for the purpose of this study.

After all, an approach to the physical point still amounts to perform the chiral and continuum limits.
In order to extrapolate the masses and (the ratio of) decay constants from the GEVP analysis to this point comprising both limits simultaneously, we have employed a simple ansatz with a linear term in $m^2_\pi$ and leading cutoff effects, in the non-perturbatively $\mathrm{O}(a)$ improved theory, proportional to $a^2$:
\begin{equation}
X(m_\pi,a)=X_0 + X_1\times m^2_\pi + X_2\times\left(\frac{a}{a_{\beta=5.3}}\right)^2 .
\label{hlmchipt}
\end{equation} 
Note that our calculation, with two lattice spacings below $0.1\,\fm$ and $(a_{\rm max}/a_{\rm min})^2\approx 1.8>1.4$ fulfills the criteria of the second level (out of three) within the quality rating regarding the continuum extrapolation by the FLAG Working Group \cite{Aoki:2016frl}. Since the datasets actually involve lattice spacings only, which both are even below $0.07\,\fm$, and the underlying action and composite fields are improved and renormalized non-perturbatively, we consider the linear description of the lattice spacing dependence in eq.~(\ref{hlmchipt}) to be adequate to estimate the systematic uncertainty related to residual discretization errors. This is also in accordance with earlier findings in the $B$ meson sector building upon almost the same collection of CLS datasets~\cite{Bernardoni:2013xba,Bernardoni:2014fva}.  

The strange and charm (valence) quark mass hopping parameters $\kappa_s$ and $\kappa_c$ were tuned at every sea quark mass (given by $\kappa_{\rm sea}$) such that $m_{D_s}(\kappa_{\rm sea};\kappa_s,\kappa_c)=m^{\rm phys}_{D_s}$. As a consequence, the formulas inspired by partially quenched heavy-light meson chiral perturbation theory (HLM$\chi$PT) \cite{SharpeQP}, which commonly serve as a guide to extrapolate lattice results of heavy-light meson masses and decay constants to the physical point, can be simplified to a leading-order (LO) chiral expression as done with our model ansatz, eq.~(\ref{hlmchipt}). As a cross-check, we have also added a next-to-leading (NLO) HLM$\chi$PT term, which has the logarithmic form $\propto m^2_\pi \log m^2_\pi$ with fit parameter $X_4$, to this ansatz. The outcome of these LO and NLO fits to the joint chiral and continuum limits, together with the fit parameters, their errors and associated $\chi^2/{\rm d.o.f.}$'s, are detailed in Table~\ref{tab:chiptfits} of the appendix. As can be read off from it, the inclusion of a NLO contribution leads to a (in most cases noticeable) deterioration of the quality of the fits, and the corresponding fitted coefficient $X_4$ is always compatible with zero, while the physical results stay consistent. We thus do not account for the discrepancy between the LO and NLO chiral extrapolations as a separate piece in the systematic error, because there are not enough data points to claim sensitivity of our data to such an NLO term, which would be a prerequisite to justify a reliable NLO analysis.

Before presenting the final results, we still have to address the effect of a possible mistuning of $\kappa_c$ and $\kappa_s$. As for $\kappa_c$, it was fixed to match $m^{\rm phys}_{D_s}=1968\,\MeV$, which together with the corresponding lattice scales in physical units, $a^{-1}=2998.89\,\MeV$ at $\beta=5.3$ and $a^{-1}=4060.23\,\MeV$ at $\beta=5.5$~\cite{Fritzsch:2012wq,HeitgerOAA}, translates into $am^{\rm phys}_{D_s}=0.656$ and $am^{\rm phys}_{D_s}=0.485$, respectively. A comparison with the numbers in the third column of Table \ref{tab:rawdata} then reveals that all data points for $am_{D_s}$ are fully compatible with $am^{\rm phys}_{D_s}$ ---~even more so upon accounting for the uncertainty on the lattice spacings from the scale setting paper~\cite{Fritzsch:2012wq}~---, except for ensemble E5. To quantify the impact of mistuning on the results from this ensemble, we have performed additional calculations at a reasonably varied hopping parameter for the valence charm quark, $\kappa_c=0.12735$ instead of $0.12724$. This yields a relative decrease by ${\rm O}(0.6\%)$ from $am_{D_s}=0.659(1)$ to $0.654(1)$, accompanied by changes in the other quantities ($af_{D_s}$, $am_{D^*_s}$, $af_{D^*_s}$, $f_{D^*_s}/f_{D_s}$) that are at most about $1/2$ of their statistical errors. As for the effect from the mistuning of the hopping parameter of the valence strange quark, $\kappa_s$ (which is largely independent of the scale setting, because it was obtained via the ratio $m^2_K/f^2_K$ in \cite{Fritzsch:2012wq}), we have studied it through additional measurements of the relevant correlators for E5 at a shifted value\footnote{Note that this shift also very well covers the updated value $\kappa_s=0.135802$ for the CLS ensemble E5, which became available only recently after including more statistics in the chiral data~\cite{DellaMorte:2017dyu}.} of $\kappa_s=0.135827$, which lies apart by three times the statistical error quoted for $\kappa_s=0.135777(17)$ in \cite{Fritzsch:2012wq}. Again, also in this case, the corresponding changes in our mesonic observables are at the same level of about $1/2$ of the respective statistical errors (or even smaller), with the exception of $f_{D^*_s}/f_{D_s}$ that slightly drops to $1.23(2)$ but within its error remains consistent with the value $f_{D^*_s}/f_{D_s}=1.25(2)$ in Table~\ref{tab:rawdata}. Therefore, since a marginal mistuning of $\kappa_c$ and $\kappa_s$ only affects the (coarser and least chiral) ensemble E5, but all in all turns out to be insignificant there, and is absent for all the other ensembles, it appears safe to neglect this contribution to the systematic error beyond the uncertainty on the lattice spacing, which of course is propagated into the overall errors quoted below.

\begin{figure}
 \centering 
 \includegraphics*{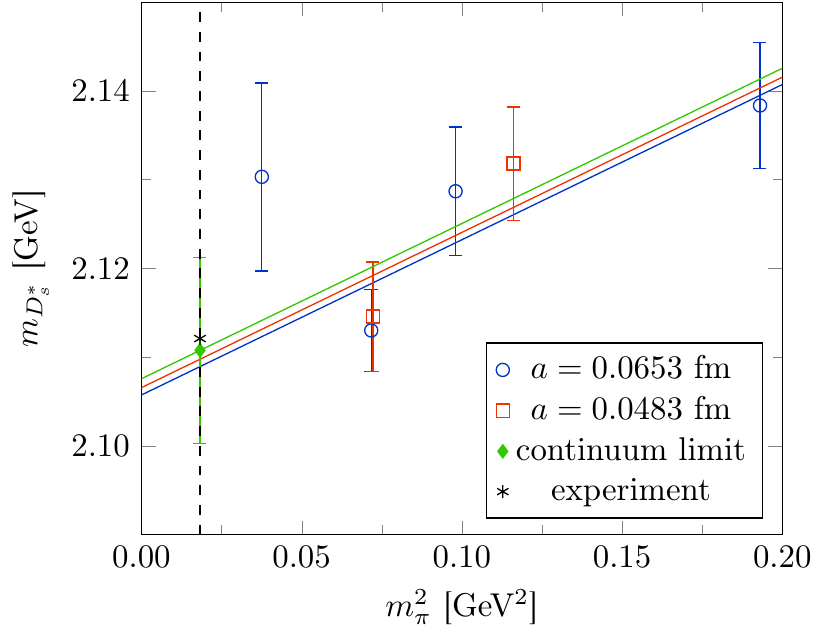}
 \caption{Joint chiral and continuum extrapolation of $m_{D^*_s}$ to the physical point, linear in $m^2_\pi$ and $a^2$.}
\label{fig2}
\end{figure}

We now come to the results of our analysis for the physical quantities under disposal. As can be inferred from Fig.~\ref{fig2}, the joint chiral and continuum limit extrapolation of the charm-strange meson mass in the vector channel, $m_{D^*_s}$, nicely reproduces its experimental value of 2.112 GeV \cite{Patrignani:2016xqp}, with cutoff effects being limited to about 0.5\% at $\beta=5.3$.
More precisely, we arrive at 
\begin{equation}
m_{D^*_s}=2.111(10)(13)\,{\rm GeV},
\end{equation}
where the first error is of statistical nature, and the second one reflects the uncertainty in the lattice spacing induced by the scale setting \cite{Fritzsch:2012wq}. 

\begin{figure}[t]
 \hspace{-2em}
 \begin{minipage}[c]{0.32\linewidth}
  \centering 
  \includegraphics*[width=1.0\linewidth]{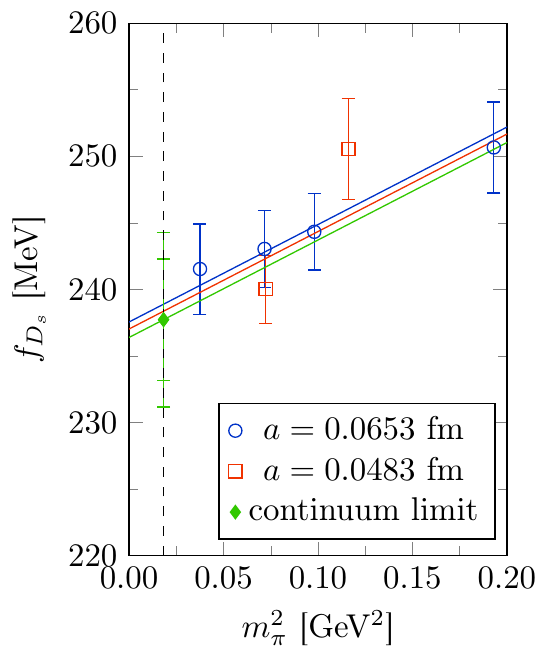}
 \end{minipage}\hspace{0.5em}
 \begin{minipage}[c]{0.32\linewidth}
  \centering 
  \includegraphics*[width=1.0\linewidth]{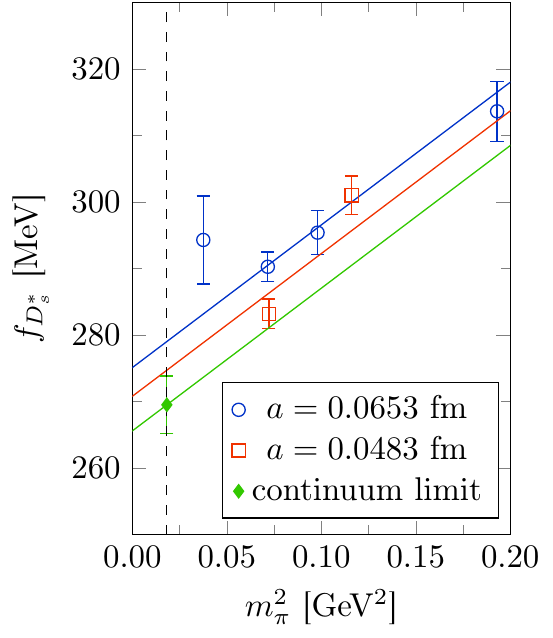}
 \end{minipage}\hspace{0.5em}
 \begin{minipage}[c]{0.32\linewidth}
  \centering 
  \includegraphics*[width=1.0\linewidth]{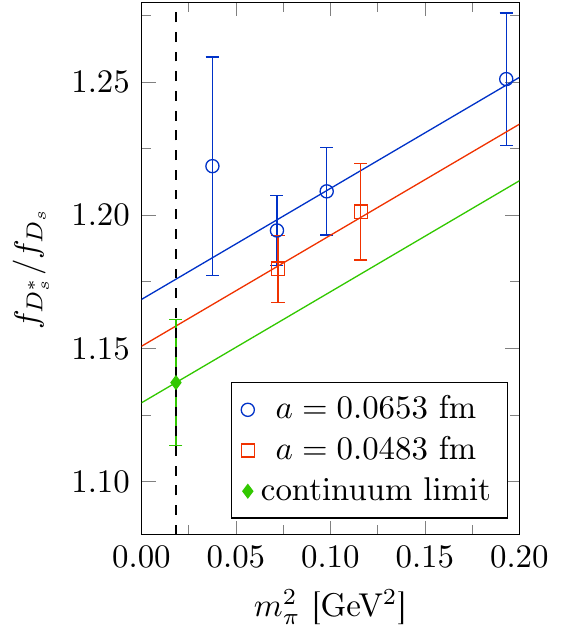}
 \end{minipage}
 \caption{Extrapolation of $f_{D_s}$ (left panel), $f_{D^*_s}$ (middle panel) and $f_{D^*_s}/f_{D_s}$ (right panel) to the physical point, through fits based on linear expressions in $m^2_\pi$ and $a^2$ as described in the text. Note that the larger error bar on the continuum limit of $f_{D_s}$ (left panel) also accounts for the uncertainty from the scale setting.}
\label{fig3}
\end{figure}

Analogous physical point extrapolations of the pseudoscalar and vector meson decay constants, $f_{D_s}$ and $f_{D^*_s}$, as well as of their ratio $f_{D^*_s}/f_{D_s}$ are displayed in Fig.~\ref{fig3} and overall show an only quite mild dependence on the pion mass and the lattice spacing. In particular, cutoff effects on $f_{D_s}$ are limited to $\sim$~1\% at $\beta=5.3$, while stronger scaling violations of the order of 7\% are observed for $f_{D^*_s}$. Hence, they also propagate with a contribution of about 6\% into the total error on the ratio $f_{D^*_s}/f_{D_s}$. We quote as our main result
\begin{equation}
f_{D^*_s}/f_{D_s}=1.14(2),
\end{equation}
where in case of this ratio the systematic error that stems from the uncertainty in lattice spacings is negligible on the level of precision here. Finally, as our estimate for $f_{D_s}$ at the physical point we obtain $f_{D_s}=238(5)(2)\,{\rm MeV}$. Again, the first error is statistical, while the second one incorporates the uncertainty from the scale setting. This value is lower by about $1.8 \sigma$ than the $N_f=2$ lattice QCD average quoted by the FLAG Working Group $f^{{\rm FLAG},N_f=2}_{D_s}=250(7)\,{\rm MeV}$ \cite{Aoki:2016frl,CarrascoZTA}\footnote{According to the FLAG criteria, there is so far only the single result \cite{CarrascoZTA} that contributes to the $N_f=2$ average of $f_{D_s}$.}. Let us emphasize, however, that the agreement in $f_{D_s}$ is satisfactory, when comparing to the outcome of the independent two-flavour computation in \cite{HeitgerOAA}, which uses almost the same CLS ensembles as in this work. There, the extraction of the $D_s$ meson decay constant follows from an expression, which combines the axial Ward identity (resp. PCAC) quark mass $m^\text{AWI}_{cs}$, cf.~eq.~(\ref{mpcac}), with the pseudoscalar-to-vacuum matrix element of the pseudoscalar density operator (the latter being obtained through a fit of the local pseudoscalar correlator $C_{P^{L}P^{L}}(t)$) and leads to results consistent with the ones reported here.

\subsection{Discussion} 

\begin{figure}[t] 
 \centering 
 \includegraphics*{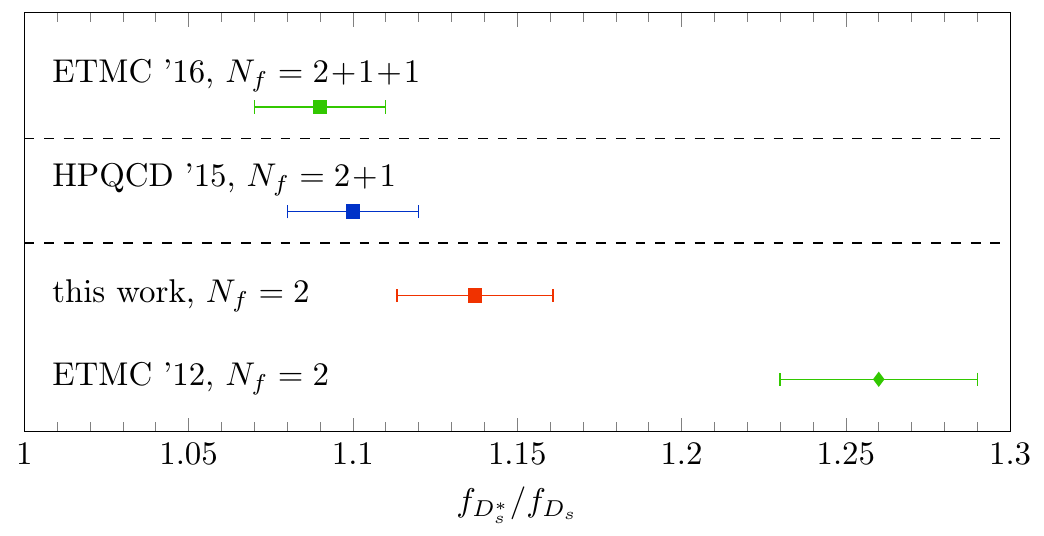}
 \caption{Collection of lattice results for $f_{D^*_s}/f_{D_s}$.}
\label{fig4}
\end{figure}

So far, there are only two lattice estimates of $f_{D^*_s}/f_{D_s}$ for $N_f=2$, namely by ETMC \cite{BecirevicTI} and us, and two other ones have been performed for $N_f=2+1$ by HPQCD \cite{DonaldSRA} and $N_f=2+1+1$ by ETMC \cite{LubiczASP}, respectively. We summarize the various results in Fig.~\ref{fig4}. In the past, owing to the apparent discrepancy between the two-flavour ETMC result of about 1.25 and the $N_f=2+1(+1)$ determinations, it was thought that $f_{D^*_s}/f_{D_s}$ could be a quantity, where a quite large quenching effect of the strange quark shows up, with an amount of $\sim$ 10\% or even more. Our finding, employing a lattice discretization of the two-flavour theory different from the ETMC calculation in \cite{BecirevicTI}, tends however to point to the conclusion that this effect is significantly less pronounced; nonetheless, the trend that less spin-breaking effects are present, if more flavours are active, still remains to be visible when placing the result of our study to the circle of the other lattice estimates in Fig.~\ref{fig4}.

\section{Conclusion}

In this paper we have reported on a two-flavour lattice QCD computation of the ratio of vector to pseudoscalar decay constants $f_{D^*_s}/f_{D_s}$, based on simulations with non-perturbatively improved Wilson fermions, satisfying non-perturbative renormalization and accounting for correlations between the two decay constants in the (statistical) error analysis. As an interesting lesson we can state that the quenching of the strange and the charm quark has an only moderate impact (of the order of $\sim $ 5\%) on a ratio, which is quite of phenomenological pertinence in quantifying the rate of spin-symmetry breaking in heavy-strange mesonic bound states. 

In a next step towards our intended study of $B_s \to D^{(*)}_s$ transitions by means of lattice QCD, a strategy inspired by the ``step scaling in mass" method of \cite{BlossierHG, AtouiZZA} will be adopted to extrapolate results to the $B$ meson region. A significant difference, however, is that in the present case of Wilson-Clover regularization we cannot use renormalization group invariant (RGI) quark masses to impose lines of constant physics for continuum limit extrapolations. Indeed, owing to the lack of knowledge of the $\mathrm{O}(a^2)$ improvement coefficient $b'_m$, which would enter in the definition of the RGI quark mass in terms of the bare (vector Ward identity) quark mass, viz. 
\begin{equation}
m^{\rm RGI}_q \equiv Z^{\rm RGI}\left(1+am_qb_m+(am_q)^2b'_m\right) ,
\end{equation}
$m^{\rm RGI}_q$ becomes negative for quark masses above the charm while setting $b'_m=0$, because $b_m<0$ \cite{Fritzsch:2010aw}. Thus our strategy is to fix a series of five values $\kappa_{h_i}$ such that 
\begin{equation}
m_{H_{is}}/m_{D_s} = (m_{B_s}/m_{D_s})^{i/6} \equiv 1.18197^{\,i/6} ,
\end{equation}
where $H_{is}$ denotes a heavy-strange meson made out of quarks with masses $\kappa_{h_i}$ and $\kappa_s$. By this we are able to access the $B$ physics region in a controlled way, which constitutes a firm basis to proceed with our program of studying there (semi-)leptonic decays and lepton-flavour universality violations through lattice QCD.

\subsection*{Acknowledgments}

This work was granted access to the HPC resources of CINES and IDRIS under the allocations 2016-x2016056808 and 2017-A0010506808 made by GENCI. In addition, it was partly supported by the grant {HE~4517/3-1} (J.~H. and M.~P.) of the Deutsche Forschungsgemeinschaft. Finally, we thank our colleagues in the CLS effort for the joint production and use of the $N_f=2$ gauge configurations.

\section*{Appendix: Numerical result details}

\renewcommand{\arraystretch}{1.2}
\setlength{\tabcolsep}{8pt}
\begin{table}[t]
\begin{center}
\begin{tabular}{ccccccc}
\hline
\hline
\toprule
id&$[t_{\rm min}, t_{\rm max}]$&$am_{D_s}$&$af_{D_s}$&$am_{D^*_s}$&$af_{D^*_s}$&$f_{D^*_s}/f_{D_s}$\\
\hline
\hline
\midrule
E5&$[10,25]$&0.659(1)&0.083(1)&0.708(2)&0.104(1)&1.25(2)\\
F6&$[10,42]$&0.657(1)&0.081(1)&0.705(2)&0.098(1)&1.21(2)\\
F7&$[10,40]$&0.655(1)&0.080(1)&0.699(2)&0.096(1)&1.19(1)\\
G8&$[11,41]$&0.656(1)&0.080(1)&0.705(4)&0.097(2)&1.22(4)\\
\hline
\midrule
N6&$[13,42]$&0.485(1)&0.061(1)&0.522(2)&0.074(1)&1.20(2)\\
O7&$[14,55]$&0.484(1)&0.059(1)&0.518(2)&0.069(1)&1.18(1)\\
\bottomrule
\hline
\hline
\end{tabular}
\end{center}
\caption{Masses and decay constants in lattice units (together with the fit intervals employed in the GEVP analysis to extract them) from the individual CLS ensembles, which enter the joint extrapolations to the chiral and continuum limits discussed in the main part of the paper. \label{tab:rawdata}}
\end{table}

In Table \ref{tab:rawdata} we list our numerical results for $m_{D_s}$, $f_{D_s}$, $m_{D_s^*}$ and $f_{D_s^*}$ in lattice units, as well as for $f_{D_s^*}/f_{D_s}$, which were obtained via the GEVP analysis of the correlation functions evaluated on the CLS gauge field ensembles contributing to this work.

\renewcommand{\arraystretch}{1.2}
\setlength{\tabcolsep}{8pt}
\begin{table}[t]
\begin{center}
\begin{tabular}{cccccc}
\hline
\hline
\toprule
fit type&fit parameter&$m_{D^*_s}~[\MeV]$&$f_{D_s}~[\MeV]$&$f_{D^*_s}~[\MeV]$&$f_{D^*_s}/f_{D_s}$\\
\hline
\hline
\midrule
LO &$X_0$&$2108(11)$           &$236(5)$               &$266(5)$             &$1.13(3)$              \\
   &$X_1$&$1.7(5)\cdot 10^{-4}$&$7.4(2.3)\cdot 10^{-5}$&$2.1(3)\cdot 10^{-4}$&$4.2(1.7)\cdot 10^{-7}$\\
   &$X_2$&$-1.8(11.0)$         &$1.1(4.9)$             &$9.4(4.5)$           &$0.04(3)$              \\
   &$\chi^2/{\rm d.o.f.}$&$2.01$&$0.88$&$2.91$&$0.275$\\
\hline
\midrule
NLO&$X_0$&$2130(29)$              &$234(11)$               &$264(15)$                &$1.19(8)$               \\
   &$X_1$&$-2.1(2.8)\cdot 10^{-3}$&$3.3(11.2)\cdot 10^{-4}$&$3.8(15.2)\cdot 10^{-4}$ &$-5.8(9.0)\cdot 10^{-6}$\\
   &$X_2$&$-5.1(12.0)$            &$1.5(5.4)$              &$9.6(4.8)$               &$0.03(3)$               \\
   &$X_4$&$1.8(2.3)\cdot 10^{-4}$ &$-2.0(8.9)\cdot 10^{-5}$&$-1.3(12.0)\cdot 10^{-5}$&$-4.9(7.1)\cdot 10^{-7}$\\
   &$\chi^2/{\rm d.o.f.}$&$2.85$&$1.31$&$4.36$&$0.283$\\
\bottomrule
\hline
\hline
\end{tabular}
\end{center}
\caption{Results of the leading-order (LO) and next-to-leading-order (NLO, i.e., also including a logarithmic HLM$\chi$PT-inspired term with fit parameter $X_4$) joint chiral and continuum extrapolations to the ansatz in eq.~(\ref{hlmchipt}). Note that the $\chi^2/{\rm d.o.f.}$'s of the underlying fits are given in the second column, too. \label{tab:chiptfits}}
\end{table}

Table \ref{tab:chiptfits} summarizes the outcome of the fits of the final results on $m_{D_s^*}$, $f_{D_s}$, $f_{D_s^*}$ and $f_{D_s^*}/f_{D_s}$ in physical units to the ansatz (\ref{hlmchipt}) for the global fit modeling their approach to the chiral and continuum limits. The next-to-leading-order (NLO) fit type also incorporates a logarithmic term from HLM$\chi$PT (proportional to $m^2_\pi \log m^2_\pi$ with fit parameter $X_4$), on top of the leading-order (LO) expression in eq.~(\ref{hlmchipt}).

\newpage


\end{document}